\documentclass[aps,a4paper,onecolumn,showpacs,showkeys,preprintnumbers,amsmath,amssymb,11pt,prd]{revtex4-1}

\usepackage[utf8x]{inputenc}

\usepackage{amsmath,amsfonts,amsthm,amssymb,epsf}
\usepackage[mathscr]{eucal}
\usepackage{graphicx}

\newtheorem*{thm}{Theorem}
\newtheorem*{dfn}{Definition}

\begin{document}

\title{Properties of Quantum Graphity at Low Temperature}

\author{Francesco Caravelli${}^{1,2}$}
\email{fcaravelli@perimeterinstitute.ca}
\author{Fotini Markopoulou${}^{1,2,3}$}
\email{fmarkopoulou@perimeterinstitute.ca}
\affiliation{${}^1$ Perimeter Institute for Theoretical Physics, \\
Waterloo, Ontario N2L 2Y5
Canada, \\and\\
 ${}^2$ University of Waterloo, Waterloo, Ontario N2L 3G1, Canada,\\
 and\\
${}^3$ Max Planck Institute for Gravitational Physics, Albert Einstein Institute,\\
Am M\"uhlenberg 1, Golm, D-14476 Golm, Germany.}

%%%%%%%%%%%%%%%%%%%%%%%%%%%%%%%%%%%%%%%%%%%%%%%%%%%%%

\begin{abstract}
We present a mapping of dynamical graphs and, in particular, the graphs used in the Quantum Graphity models for emergent geometry, to an Ising hamiltonian on the 
line graph of a complete graph with a fixed number of vertices.  We use this method to 
 study the properties of Quantum Graphity models at low temperature in the limit in which the valence coupling constant 
of the model is much greater than the coupling constants of the loop terms. Using mean field theory we find that an order parameter for the model is the average
valence of the graph.  We calculate the equilibrium distribution for the valence as an implicit function of the temperature. In the approximation
in which the temperature is low, we find the first two Taylor coefficients of the valence in the temperature expansion. A discussion of the susceptibility function and a
generalization of the model are given in the end. 
\end{abstract}

\pacs{04.60.Pp , 04.60.-m}

\maketitle
\newpage

%%%%%%%%%%%%%%%%%%%%%%%%%%%%%%%%%%%%%%%%%%%%%%%%%%%%%

\section{Introduction}

It is commonly agreed that at high spacetime curvatures, when the quantum effects of the gravitational field become significant, General Relativity needs to be replaced by a quantum theory of gravity.  In spite of 
progress in several directions, finding this new theory has proven a challenging problem for several decades.  
Current research in the field is paying substantial attention to the  numerous indications that gravity may only be {emergent}, meaning that it is a collective, or thermodynamical, description of microscopic physics in 
which we do not encounter geometric or gravitational degree of freedom.  An analogy to illustrate this point 
of view is fluid dynamics and the transition from from thermodynamics to the kinetic theory.  What we currently know is the low energy theory, the analogue of fluid dynamics.  We are looking for the microscopic theory, 
the analogue of the quantum molecular dynamics.   Just as there are no waves in the molecular theory,  we may not find geometric degree of freedom in the fundamental theory.  
 Not surprisingly, this significant shift in perspective opens up new routes that may take us out of the old problems. 

The emergent viewpoint amounts to treating quantum gravity as a problem in statistical physics.  A powerful set of methods in statistical physics involve the use of lattice-based models, such as the Ising model for 
ferromagnetism, the Hubbard model for the conductor/insulator transition, etc.  Such methods are starting to be introduced in quantum gravity.  Examples are G. Volovik's work on emergent Lorentz invariance at the Fermi 
point \cite{GV}, X.-G.\ Wen's work on emergent matter and gravitons from a bosonic spin system \cite{Wen}, the emergence of a Lorentzian metric and aspects of gravitation such as Hawking radiation in analog models of 
gravity \cite{analog}, as well as long-standing approaches such as matrix models \cite{MM}, and more radical formulations of geometry in terms of information \cite{Llo,FMKQI}. 

It is natural for the lattice in the lattice system to play the role of (a primitive form of) geometry.  Now, General Relativity is a background independent theory, by which we mean that the geometry of spacetime is fully 
dynamical. 
By analogy, we expect 
 that the use of a fixed lattice is inappropriate and one instead needs models on a {\em dynamical} lattice.  While for this reason desirable, dynamical lattices raise difficult technical problems that have not been previously 
addressed in the field of statistical physics.  The present article is concerned with exactly this problem and presents a method that deals with dynamical lattices in certain situations.  

In previous work, we introduced Quantum Graphity, a background independent model of spacetime in which time is an external parameter and space
is described by a relational theory based on graphs \cite{graphity1}. The idea is to represent locality by the adjacencies of a dynamical graph on which the diffeomorphism group is replaced, in the high energy phase, by the 
symmetric group on the complete graph
which breaks down to its subgraphs at lower energies. These kind of models are sometimes called \textit{event-symmetric}\cite{events}).  The principle of event symmetry refers to the replacement, at high energies, of 
the diffeomorphism group with the group of permutations of events in spacetime. In the context of Quantum Graphity, however, the event symmetry is only spatial, in the sense that at high energy the graph is complete and every 
vertex of the graph is at distance one from each other.  If the dynamics is such that the system settles into a minimum energy subgraph that exhibits geometric symmetries, for instance, a discrete version of flat space in low 
dimension, we say that geometry emerges in that phase. 
  In \cite{graphity1} was shown that desired symmetric lattices, i.e., discrete 2d FRW, are stable local minima under certain choices of parameters in the hamiltonian (when the effect of the matter on the lattice is neglected).  
  Following this work, in
   \cite{graphity2}, we used the same concept of locality in terms of a dynamical lattice, but with a new type of matter that interacts non-linearly with the geometry, a precursor of gravity, and initiated a study of the quantum 
properties of that system.  

In the present article, we return to the technical issues of spin systems on dynamical lattices and
we introduce a method to deal with a theory of dynamical graphs on $N$ vertices.  Such graphs are subgraphs of $\mathscr K_N$, the complete graph on $N$ vertices.  We show how, 
by transforming $\mathscr K_N$ to its line graph, the theory can be approximated by an Ising model on the line graph of a complete graph.  We then use this to 
 study the low energy properties of the Quantum Graphity model in \cite{graphity1}.   
Using mean field theory we calculate the average valence of the graph at low temperature and we evaluate the first corrections due to the presence of 3-loops. 

The paper is organized as follows. 
In section II, we review the Quantum Graphity model \cite{graphity1} with no matter.  In section III, we define the line graph derived from a generic graph and summarize its properties. In section IV, we show how, in a certain reasonable 
approximation, the 
hamiltonian of Quantum Graphity can be recast as an Ising hamiltonian on the line graph of a complete graph. In section V, we calculate the corrections due to loops at low temperature and describe, in this framework, the 
behavior of the correlation
function in mean field theory. Conclusions follow in Section VI.

%%%%%%%%%%%%%%%%%%%%%%%%%%%%%%%%%%%%%%%%%%%%%%%%%%%%%

\section{Quantum Graphity}
Let us briefly introduce the Quantum Graphity model \cite{graphity1}. As the name {\em graphity} implies, Quantum Graphity is a model for a quantum theory of gravity in which the fundamental microstates are dynamical  
graphs, postulated to describe relational physics at Planckian energies.  There is no notion of geometry or quantum geometry at high energy, instead, geometry emerges as the system cools down and away from the Planckian regime.  
The microstates live in a Hilbert space on the complete graph ${\mathscr K}_N$ of $N$ vertices, given by
$$
\mathscr H = \bigotimes^{\frac{N(N-1)}{2}}_{ij} \mathscr H^e_{ij} \bigotimes^{N}_j \mathscr H^v_{j},
$$
where $e_{ij}$ represents the edge of the graph $K_N$ between the $i$ and $j$ vertices, while $\mathscr H^e_{ij}$ and $\mathscr H^v_j$ are the Hilbert spaces associated with edges and vertices respectively. 
In particular, the Hilbert space associated with an edge between vertex $i$ and $j$ is the two-level state space:
\begin{equation}
\mathscr H^e_{ij}=\text{span}\{ |0\rangle,|1\rangle\}.
\label{hilberts}
\end{equation}
The two states 1,0 in (\ref{hilberts}) are interpreted as the edge being \textit{on} or \textit{off} respectively.
This choice means that basis states in the Hilbert space of the edges represent  subgraphs  $\mathscr G_{{\mathscr K}_N} $ of the complete graph ${\mathscr K}_N$. A generic state in the Hilbert space of the 
edges is a superposition of such subgraphs:
$$
|\psi\rangle=\sum c_{t} \mathscr G_{K_N; t}.
$$
In the full model of \cite{graphity1}, extra degrees of freedom are assigned to the on states:
\begin{equation}
\mathscr H^e_{ij} =\text{span}\{|0\rangle_{ij},|1_1\rangle_{ij},|1_2\rangle_{ij},|1_3\rangle_{ij}\}.
\label{fermhilb} 
\end{equation}
In \cite{graphity1}, and in the present work, there are no degrees of freedom associated with the vertices and hence we ignore ${\mathscr H}^v$.  

Let us focus now on the state space (\ref{hilberts}). On the 
%fermionic 
Fock space of the edges we can introduce
the ladder operators $\hat a$ and $\hat a^\dagger$, with the usual action:
$$
\hat a^{\dagger}_{ij}|1\rangle_{ij}=\hat a_{ij} |0\rangle_{ij}=0\ ,\ \hat a|1\rangle_{ij}=|0\rangle_{ij}.
$$
%and which satisfy the anticommutator relations:
%$$\{a_{ij},a^\dagger_{ij}\}=1.$$
Dynamics in Quantum Graphity is given by a hamiltonian acting on the graph states of the form\cite{graphity1}: 
\begin{equation}
 \widehat H = \widehat H_V + \widehat H_L + interaction\ terms,
\label{hamiltonian}
\end{equation}
where $\widehat H_V$ keeps track of how many \textit{on} edges are attached to a single vertex,
and $\widehat H_L$ counts closed paths in the graph.
The interaction term will not be used in the following, but in a generic model these terms produce Alexander moves on the graph. 

 In more detail, the term $\widehat H_V$ is usually chosen to be of the form:
\begin{equation}
\widehat H_V = g_V \sum_{j} e^{p \left(v_0 \widehat 1-\sum_i \widehat N_{ij}\right)^2},
\label{vertext}
\end{equation}
where the indices $i,j=1,...,N$ enumerate vertices, $\widehat N_{ij}=\widehat a^{\dagger}_{ij} \widehat a_{ij}$ is the usual number operator, and $g_V$ and $p$ are free couplings that we assume to be positive. 
The purpose of this term is to ensure that at 
low energies the system has a (low temperature) phase in which the average vertex valence (i.e.\ \textit{on} edges attached to a vertex) is $v_0$. Later on in the paper we will show that, at least in 
the mean field theory  
approximation, this is indeed the case. The term $\widehat H_L$ is given by
\begin{equation}
\widehat H_L = - g_L \sum_i \sum_L \frac{r^L}{L!} \widehat P(i,L),
\label{patht}
\end{equation}
where $g_L$ and $r$ are couplings assumed to be positive. The operator $\widehat P(i,L)$ counts the number of non-retracing paths of length L based at the vertex $i$. This operator is related to the trace of the 
adjacency matrix in the original
model. We will build this operator in another way in the following.
For $r\leq1$, so that higher length loops contribute less than short length loops, this term is semi-local. The $L!$ comes from the expansion of the exponential of the loop-path operator\cite{graphity1}.
It was shown in \cite{graphity1} that  $r$ determines the length on which loop size is peaked at low energies. 

In what follows we introduce a new method to analyze dynamical lattices, by transforming ${\mathscr K}_N$ to its {\em line graph} which we define in the next section.  The {\em on/off} edges of ${\mathscr K}_N$ will 
become Ising spins on the {\em fixed}  line graph, so that standard methods of statistical physics can be used.

%%%%%%%%%%%%%%%%%%%%%%%%%%%%%%%%%%%%%%%%%%%%%%%%%%%%%

\section{Graphs and Line Graphs}

\begin{figure}
\centering
\includegraphics[scale=0.5]{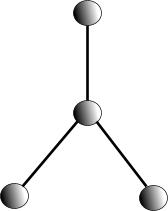} 
\caption{The simplest example of element of the Beineke classification of a minimal graph that is not a line graph of any other one. This means that in general there is not a one-one correspondence between a 
graphs and \emph{line-duals}. }
\label{beineke}
\end{figure}

We start by defining line graphs.
 Let  $G=(V,E)$ denote a graph with vertex set $V=\{v_1,v_2,...\}$ and edge set  $E=\{e_1,e_2,...\}$.
The line graph $\mathscr L(G)=(\widetilde{V},\widetilde{E})$ is the graph of the adjacencies of $G$, containing information on the connectivity of the original graph.
Each vertex $\tilde v \in {\widetilde V}({\mathscr L}(G))$ corresponds to an edge $e \in E(G)$. Two vertices $\tilde v_1$ and $\tilde v_2$ in $\widetilde V(\mathscr L(G))$ are adjacent if and only if the edges in G 
corresponding to $\tilde v_1$ and $\tilde v_2$ share a vertex. The correspondence between $G$ and $\mathscr L(G)$ is not one to one. From a given graph $G$ we can construct only one $\mathscr L(G)$ but it is not true that 
any graph is a line graph. In fact, according to the Beineke classification, there are 9 non-minimal graphs that are not line graphs of another graph and each graph containing them is not a line graph\cite{beineke}.
The simplest example of a line graph is depicted in Fig. \ref{beineke}.

Given a graph $G$, we can construct its line graph using the following procedure:
\begin{enumerate}
\item  Enumerate the vertices of $G$.
\item Enumerate the edges of $G$ with a fixed prescription (see example below) and put a blob on them.
\item If two edges share a vertex, draw a bold line between them.
\item Remove $G$ and its enumeration.
\end{enumerate}
What is left is the line graph of $G$  where the blobs represent its vertices.

 Let us now introduce some useful quantities:

\begin{dfn} (\textit{Kirchhoff matrix})
 Let $G$ be a generic graph, $V=\{v_1,v_2,\cdots,v_n\}$ be the set of vertices of $G$ and $E=\{e_1,e_2,\cdots,e_p\}$ be the set of
edges of $G$. Let us define the matrix $P$ of size $n \times p$ with entries $P_{i\beta}$, where $i$ is an integer between $1$ and $n$ on the set of vertices and $\beta$ is an integer between $1$ and $p$ on 
the set of edges, such that:
\begin{equation}
  P_{i\beta} = \left\{
    \begin{array}{rl}
      1 & \text{if the edge $\beta$ has an vertex on the vertex $i$} ,\\
      0 & \text{otherwise }.
    \end{array} \right.
\label{pij}
\end{equation}
The Kirchhoff matrix $K$ is the $p\times p$ matrix built from $P$, such that:
\begin{equation}
K=P^t\ P,
\label{kirchmatr}
\end{equation}
$P^t$ representing the transpose of $P$.
\end{dfn}

% The Kirchhoff matrix has the property that:
% \begin{equation}
% \sum_{\alpha} K_{\beta \alpha} K_{\alpha \gamma}= K_{\beta \gamma}.
% \label{kircprop} 
% \end{equation}
A well-known theorem now gives the incidence matrix of the line graph $\mathscr L(G)$:

\begin{thm}{ {Let $G$ be a graph with p edges and n vertices and let $\mathscr L(G)$ be its line graph. Then the matrix:}
\begin{equation}
J = K - 2\ \textbf{I},
\label{kirchhoff}
\end{equation}
where $\textbf{I}$ is the $p\times p$ identity matrix, is the incidence matrix of $\mathscr L(G)$}.
\end{thm}

In the next section we will show how the graphity hamiltonian can be recast on the line graph using (\ref{pij}) and (\ref{kirchhoff}).

%%%%%%%%%%%%%%%%%%%%%%%%%%%%%%%%%%%%%%%%%%%%%%%%%%%%%

\section{The line graph representation}

Since in the hamiltonian (\ref{hamiltonian}) we neglect the terms in which vertices are interacting because we assume there are no degrees of freedom on them, one could expect that it
can be rewritten only in terms of the connectivity of the graph. To carry out such a reformulation, let us expand the first term in (\ref{hamiltonian})
for small values of the parameter $p$:
\begin{eqnarray}
 \widehat H_V&=& g_V \sum_{i} \widehat 1+ p g_V \sum_i \left(v_0-\sum_j \widehat N_{ij}\right)^2 + {\mathscr O}(p^2) \nonumber \\
& = &g_V\left(1+v_0^2p\right) \sum_i \widehat 1 + p g_V \sum_{ijk} \widehat N_{ij}\widehat N_{jk}-2 g_V p v_0 \sum_{ij} \widehat N_{ij}+{\mathscr O}(p^2).
\label{exp}
\end{eqnarray}
As we will see later, such an expansion does not modify the properties of the model at low temperature. The first term in (\ref{exp}) is an energy shift and, for what is to come, can be neglected.
We should now be able to recognize some particular terms in the expansion (\ref{exp}). The third term is proportional to the operator
$\sum_{ij} \widehat N_{ij} $.
It is the sum over all the edges of the graph, zero or not, of the number operator. We will change the  notation to 
$$\sum_{ij} \widehat N_{ij} \rightarrow 2 \sum_\beta \widehat N_\beta, $$
where $\beta$, as in the previous section, runs from $1$ to $N(N-1)/2$ and labels the edges of ${\mathscr K}_N$ or, equivalently in what follows,
the vertices of its line graph. 

\begin{figure}
\centering
\includegraphics[scale=0.7]{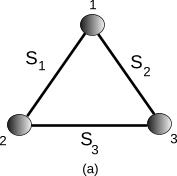} 
\includegraphics[scale=0.7]{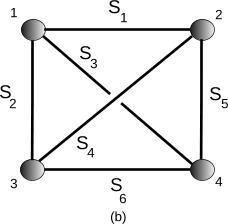}
\includegraphics[scale=0.7]{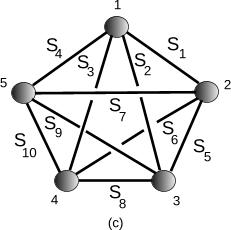}
\caption{Three examples of complete graphs labeled according to the prescription (\ref{prescr}).}
\label{completes}
\end{figure}

To rewrite the second term in (\ref{exp}), we need the matrix $P_{i\beta}$ of (\ref{pij}) in this context. This matrix maps the graph to its line graph, as we will see. We first fix a prescription to label
edges. Let $\mathscr K_N$ be the complete graph of $N$ vertices. Let $\mathscr I$ be any enumeration of $V(\mathscr K_N)$, $i\in \mathscr I=1,...,N$. 
We identify edges by their endpoint vertices $(i,j)$, with $i,j \in \mathscr I$. A {\em labeling} $S_\beta$, $\beta \in \mathscr B=\{1,\cdots,N(N-1)/2\}$ is an enumeration 
of $\widetilde E(\mathscr L(\mathscr K_N))$, according to the
following prescription:
\begin{eqnarray}
& S_1,\cdots,S_{N-1}\ \text{label the edges connecting the vertices}\ \{(1,2),\cdots,(1,N)\}; \nonumber \\
& S_N,\cdots,S_{2(N-1)}\ \text{label the edges connecting the vertices}\ \{(2,3),\cdots,(2,N)\}; \nonumber \\
& \vdots \nonumber \\
& S_{N(N-1)/2}\ \text{labels the edge connecting the vertices}\ (N-1,N).
\label{prescr}
\end{eqnarray}
Using this prescription it is easy to see that the matrix $P_{i\beta}$ introduced in (\ref{pij}), for the complete graph $\mathscr K_N$, has the simple (recursive) form:
\begin{equation}
  P^N = \left(
    \begin{array}{cc}
      \vec V_{N-1} & \vec 0 \\
      \textbf{I}^{N-1}_{b' c'} & P^{N-1}_{a' \alpha'}  \\
     \end{array} \right),
\label{Pmatr}
\end{equation}
where $\vec V_{N-1}$ is a row vector of length $N-1$, $\textbf{I}^{N-1}$ is the identity matrix of size $(N-1) \times (N-1)$ and $\vec 0$ represents a null row vector 
of length $N(N-1)/2-(N-1)$. The indices $\{a',b',c'\}$ and $\alpha'$ run from $1$ to $N-1$ and $1$ to $(N-1)(N-2)/2$ respectively. As an example, for the graphs of Fig.\ \ref{completes} the $P$ matrices
are:
\begin{equation}
  P^3 = \left(
    \begin{array}{ccc}
      1 & 1 & 0 \\
      1 & 0 & 1  \\
      0 & 1 & 1  \\
     \end{array} \right),
\label{Pmatr3}
\end{equation}
\begin{equation}
  P^4 = \left(
    \begin{array}{cccccc}
      1 & 1 & 1 & 0 & 0 & 0 \\
      1 & 0 & 0 & 1 & 1 & 0 \\
      0 & 1 & 0 & 1 & 0 & 1 \\
      0 & 0 & 1 & 0 & 1 & 1
     \end{array} \right),
\label{Pmatr4}
\end{equation}
\begin{equation}
  P^5 = \left(
    \begin{array}{cccccccccc}
      1 & 1 & 1 & 1 & 0 & 0 & 0 & 0 & 0 & 0 \\
      1 & 0 & 0 & 0 & 1 & 1 & 1 & 0 & 0 & 0   \\
      0 & 1 & 0 & 0 & 1 & 0 & 0 & 1 & 1 & 0  \\
      0 & 0 & 1 & 0 & 0 & 1 & 0 & 1 & 0 & 1 \\
      0 & 0 & 0 & 1 & 0 & 0 & 1 & 0 & 1 & 1
     \end{array} \right),
\label{Pmatr5}
\end{equation}
for (a), (b) and (c) respectively.

It is easy to see that two edges $\alpha$ and $\beta$ have a common vertex if and only if we have:
\begin{equation}
\sum_{i \in \mathscr I} {P^t}_{\beta i} P_{i \alpha}=c\neq0,
\end{equation}
where $\phantom c^t$ is the transposition operation. By construction, $c$ can take the following values only:
\begin{equation}
  c = \left\{
    \begin{array}{rl}
      2 & \text{if}\ \alpha=\beta ,\\
      1 & \text{if}\ \alpha\neq\beta\ \text{and}\ \alpha\ \text{and}\ \beta\ \text{have a common vertex}  ,\\
      0 & \text{if}\ \alpha\neq\beta\ \text{and}\ \alpha\ \text{and}\ \beta\ \text{do not have a common vertex.}
    \end{array} \right.
\label{cvalue}
\end{equation}
In particular, for $N=4$, $K$ is given by:
\begin{equation}
  K^4 = \left(
    \begin{array}{cccccc}
      2 & 1 & 1 & 1 & 1 & 0 \\
      1 & 2 & 1 & 0 & 0 & 1 \\
      1 & 1 & 2 & 0 & 1 & 1 \\
      1 & 0 & 0 & 2 & 1 & 1 \\
      1 & 0 & 0 & 1 & 2 & 1 \\
      0 & 1 & 1 & 1 & 1 & 2 \\
     \end{array} \right),
\label{kmatr4}
\end{equation}
Using now the matrix $P_{i \beta}$ just introduced, we want to construct generic \textit{n-string matrices} as composition of $n$ edges of the graph, thus a \emph{path} on the graph. 
The $n$-string matrices will be needed both for the non-retracing loop term and the 2-edge interaction term 
 in equation (\ref{exp}). The quantity:
\begin{equation}
 K^i_{\alpha \beta}=P^t_{\alpha i} P_{i \beta}
\end{equation}
is the definition of the Kirchhoff matrix of equation (\ref{kirchhoff}) if we sum over the index $i$.
From $K^i_{\alpha \beta}$ we can construct strings of $P$'s of the form
$$
Q^{i_1 \cdots i_n}_{\alpha_1 \cdots \alpha_{n+1}}
=K^{i_1}_{\alpha_1 \alpha_2} K^{i_2}_{\alpha_2 \alpha_3}\cdots  K^{i_n}_{\alpha_n \alpha_{n+1}}, 
$$
that we call \textit{string matrices} of \textit{n}th order. These string matrices represent paths through vertices $i_1 \cdots i_n$ and they are zero
unless the edges corresponding to $\alpha_1 \cdots \alpha_{n+1}$ are in the correct order, that means, they represent an actual path on the graph. For instance, the number of paths of length 
2 on the complete graph is given by
\begin{equation}
\# 2\mbox{-strings}= \sum_{\alpha\neq \beta \in \mathscr B}\ \ \sum_{i \in \mathscr I} Q^{i}_{\alpha \beta} = \sum_{\alpha\neq \beta \in \mathscr B} K_{\alpha \beta}, 
\label{interacting1} 
\end{equation}
or, equivalently, we can use (\ref{pij}) and rewrite (\ref{interacting1}) as
\begin{equation}
\#{\mbox{2-strings}}= \sum_{\alpha \beta \in \mathscr B}\ \  \sum_{i \in \mathscr I} Q^{i}_{\alpha \beta} = \sum_{\alpha \beta \in \mathscr B}\left(K_{\alpha \beta}-2\ \textbf{I}_{\alpha \beta}\right).
\label{interacting2} 
\end{equation}
The subtraction of twice the identity in (\ref{interacting2}) is the same as the subtraction of the self-energy of each edge. We  now clearly see that this matrix is precisely the incidence matrix of the
line graph of $\mathscr K_N$ introduced in (\ref{kirchhoff}), with $\alpha,\beta \in \widetilde V(\mathscr L(\mathscr K_N))$.

So far we have dealt with the complete graph only.   We wish to extend this formalism to a dynamical graph.  In order to do that we return to the Hilbert space formulation of the graph with \emph{on/off} edges.
 Recall that any graph on $N$ vertices is a subgraph of the complete graph $\mathscr K_N$, with some edges  \emph{off}. Thus, since we can always map a graph on a 
complete graph, we can count paths on any graph by modifying
(\ref{interacting2}) so that it  counts paths of only {\em on} edges on the corresponding complete graph. To do so, we introduce in the sum the number operators $\widehat N_\beta$ in the following way:
\begin{eqnarray}
\# \mbox{2-strings}&
=& \sum_{\alpha \beta \in \mathscr B}
\left(K_{\alpha \beta}-2\ I_{\alpha \beta}\right) \widehat N_\alpha \widehat N_\beta\nonumber \\
&= &\sum_{\alpha \beta \in \mathscr B}J_{\alpha \beta}\ \widehat N_\alpha \widehat N_\beta.
\label{2edge} 
\end{eqnarray}
This term does not contribute if any of the two edges $\alpha,\beta$ is off. It is easy to see that this term of the hamiltonian is an Ising interaction. The important difference between these two hamiltonians is 
that in our case the spin system is on
the line graph a complete graph $\mathscr K_N$. 

By extension of the above, we are now able to construct a generic \textit{path operator} out of $K^i_{\alpha \beta}$'s. We define
\begin{eqnarray}
\widehat P(n) 
&:= &\sum_{\mathscr Q} \sum_{\alpha_1 \cdots \alpha_n} K^{i_1}_{\alpha_1 \alpha_2}K^{i_2}_{\alpha_2 \alpha_3}\cdots K^{i_{n-1}}_{\alpha_{n-1} \alpha_n}
 \ \widehat N_{\alpha_1} \cdots \widehat N_{\alpha_n}\nonumber\\
&  = &\sum_{\mathscr Q} \sum_{\alpha_1 \cdots \alpha_n} Q^{i_1 \cdots i_{n-1}}_{\alpha_1 \cdots \alpha_{n}} \ \widehat N_{\alpha_1} \cdots \widehat N_{\alpha_n},
\label{pathop} 
\end{eqnarray}
where the set $\mathscr Q$ is
\begin{equation}
  \mathscr Q = \left\{
    \begin{array}{l}
      i_1 \neq \cdots \neq i_{n-1} \in \mathscr I  \text{ for non-retracing paths},\\
      i_1,\dots,i_{n-1} \in \mathscr I  \text{ for retracing paths}.\\
    \end{array} \right.
\label{Qset}
\end{equation}
It is easy to see that it counts the number of paths of length $n$ in the graph, that is why we call the $Q$'s \emph{string matrices}. Note that  $Q^{i_1 \cdots i_{n-1}}_{\alpha_1 \cdots \alpha_{n}}$ can 
take  values $0$ and $1$ only because it is a product of $0$'s and $1$'s. This string matrix is \textit{not} the matrix multiplication of the Kirchhoff matrices: it only reduces to matrix multiplication for retracing paths 
where we sum over all possible vertices. 

In the following, we will denote the two sets in (\ref{Qset}) as $\mathscr Q ^{r}$ and $\mathscr Q^{nr}$ for the retracing and non retracing cases respectively; moreover, we may explicitly show the indices on which we are 
doing the sum as $\mathscr Q^{r/nr}(i_{b(j)})$.
In order to count loops, we just need to impose $\alpha_1=\alpha_{n}$:
\begin{eqnarray}
& \underbrace{P_{\alpha_1 i_1} P_{i_1 \alpha_2}} \underbrace{P_{\alpha_2 i_2} P_{i_2 \alpha_3}} \underbrace{P_{\alpha_3 i_3} P_{i_3 \alpha_1}}. \nonumber \\ 
&\ \ K^{i_1}_{\alpha_1 \alpha_2}\ \ \ \ \ K^{i_2}_{\alpha_2 \alpha_3}\ \ \ \ \ \ K^{i_3}_{\alpha_3 \alpha_1} \nonumber
\end{eqnarray}

Thus we have discovered that, when there are no degrees of freedom on the vertices of the graph and we neglect the interaction terms, we can recast the Quantum Graphity hamiltonian on the line 
graph ${\mathscr L}({\mathscr K}_N)$ representation in the weak coupling regime at finite $N$.

We end this section with two properties of the \textit{n}th-order string matrices. Let us define:
\begin{equation}
\widetilde Q^{r/nr}_{\alpha_1 \cdots \alpha_n} = \sum_{\mathscr Q^{r/nr}} Q^{i_1 \cdots i_{n-1}}_{\alpha_1 \cdots \alpha_n}.
\label{Qtilde}
\end{equation}
The following  properties of the sum of these string matrices on complete graphs will be required next:\\
\textbf{Property 1:} Let $\mathscr{G}=\mathscr{K}_N$. Then, for a loop of $n$ edges:
\begin{equation}
\sum_{\alpha_1\neq\alpha_2\neq\cdots\neq\alpha_{n}} \widetilde Q^{nr}_{\alpha_1 \cdots \cdots \alpha_{L} \alpha_1} = N(N-1)\cdots(N-L)
\label{property1a} 
\end{equation}
and
\begin{equation}
\sum_{\alpha_1\neq\alpha_2\neq\cdots\neq\alpha_{n}} \widetilde Q^{r}_{\alpha_1 \cdots \cdots \alpha_{L} \alpha_1} = N^L.
\label{property1b}
\end{equation}
{\it Proof.} These two facts follow trivially if we note that the equations (\ref{property1a}) and (\ref{property1b})  count the number of retracing and non-retracing paths of length $L$ on the 
complete graph respectively.
\\\ \\
\textbf{Property 2:} Let $\mathscr{G}=\mathscr{K}_N$. Then, for a loop of $n$ edges, and for $L\ge4$, we have:
\begin{eqnarray}
& \sum_{\alpha_3\neq\alpha_4\neq\cdots\neq\alpha_{n}} \widetilde Q^{nr}_{\alpha_1 \cdots \cdots \alpha_{L} \alpha_1} = (N-3)\cdots (N-3-(L-4)) K_{\alpha_1 \alpha_2},\\ \nonumber
\label{property2a}
\end{eqnarray}
while, for $L=3$:
\begin{eqnarray}
& \sum_{\alpha_3} \widetilde Q^{nr}_{\alpha_1 \alpha_2 \alpha_{3} \alpha_1} =K_{\alpha_1 \alpha_2}, \\ \nonumber
\label{property2b}
\end{eqnarray}
if $\alpha_1\neq\alpha_2\neq\alpha_3\cdots\neq\alpha_n$.\\\ \\ 

{\it Proof.} Note that $\sum_{\alpha_3\neq\alpha_4\neq\cdots\neq\alpha_{n}} \widetilde Q^{nr}_{\alpha_1 \cdots \cdots \alpha_{n} \alpha_1}$ is the number
of \textit{non-retracing} loops of length $L$ on the complete graph $\mathscr K_N$ which pass by the edges $\alpha_1$ and $\alpha_2$. 
Now, it is easy to see that if the edges $\alpha_1$ and $\alpha_2$ do not share a link this quantity is zero. Also note that, by the symmetry of the complete graph, the number of non-retracing 
loops based on two neighboring edges must be the same for each pair of edges $\alpha_{j_1}$,$\alpha_{j_2}$ sharing a node. Since the matrix $K_{\alpha_{j_1}\alpha_{j_2}}$ takes values 
$1$ or $0$ depending on whether the edges $\alpha_{j_1}$,$\alpha_{j_2}$ are neighbors or not,   $\sum_{\alpha_3\neq\alpha_4\neq\cdots\neq\alpha_{n}} \widetilde Q^{nr}_{\alpha_1 \cdots \cdots \alpha_{L} \alpha_1}$ 
must be proportional to the matrix $K_{\alpha_1 \alpha_2}$. In order to evaluate the proportionality constant, let us note that each loop is weighed by a factor of 1 because n-string matrices take values $1$ or $0$ only.
The combinatorial quantity $(N-3)\cdots (N-3-(L-4))$ is then the number of non-retracing loops of length $L$ passing from two consecutive fixed edges on the complete graph of $N$ vertices, as can be easily checked. The special case (\ref{property2b}) follows from the fact
that if we fix two edges there is only one edge which closes the 3-loop. 
\\\ \\
Note that, for $N\gg L$, we have: 
\begin{equation}
 \sum_{\alpha_3\neq\alpha_4\neq\cdots\neq\alpha_{n}} \widetilde Q^{nr}_{\alpha_1 \cdots \cdots \alpha_{L} \alpha_1}\approx \sum_{\alpha_3\neq\alpha_4\neq\cdots\neq\alpha_{n}} \widetilde Q^{r}_{\alpha_1 \cdots \cdots \alpha_{L} \alpha_1}=N^L.
\end{equation}
We can now collect the results of this section to write the hamiltonian (\ref{exp}) as 
\begin{equation}
 \widehat H=
 A \sum_{\alpha, \beta \in \mathscr B} J_{\alpha \beta} \widehat N_\alpha \widehat N_\beta 
 - B \sum_{\alpha \in \mathscr B} \widehat N_\alpha 
 - C \sum_{\alpha\neq\beta\neq\gamma \in \mathscr B}
 \widetilde Q^{nr}_{\alpha,\beta,\gamma}\ \widehat N_\alpha \widehat N_\beta \widehat N_\gamma,
\label{Ising}
\end{equation}
where
\begin{eqnarray}
 A& = &p\ g_V, \nonumber \\
 B &=& 2\ g_V\ p\ v_0, \nonumber \\
 C &=& g_L\ \frac{r^3}{6},
\label{couplings}
\end{eqnarray}
and neglecting higher order loop terms. 

%%%%%%%%%%%%%%%%%%%%%%%%%%%%%%%%%%%%%%%%%%%%%%%%%%%%%

\section{Mean field theory approximation and low temperature expansion}
%%%%%%%%%%%%%%%%%%%%%%%%%%%%%%%%%%%%%%%%%%%%%%%%%%%%%

Having rewritten the hamiltonian in an Ising fashion, we now can approach the problem of finding a graph observable and its equilibrium distribution using mean 
field theory. As we will see, the  natural graph observable to consider  is the average valence of the graph. We will assume that the system is at equilibrium and we neglect the interaction terms.
%Moreover we suppose that this external temperature change is adiabatic. We 
%are then  in a phase in which the interaction terms can be neglected.
In this case, it is straightforward to use mean field theory analysis \cite{Parisi}. In what follows, we assume units in which the Boltzmann constant $k_B=1$. 

We start by replacing the number operators $\widehat N_{\alpha}$ with semi-classical analogs,
imposing that their expectation value must lie in the interval $I=[0,1]$:
$$ \widehat N_{\beta} \rightarrow \langle\widehat N_\beta\rangle_P=m_\beta, $$
where $P$ is a probability measure of the following form:
\begin{equation}
P(m_\beta) = m_\beta \delta_{1,m_\beta}+ (1-m_\beta) \delta_{0,m_\beta}.
 \label{probmeas}
\end{equation}
It is easy to see that this probability distribution forces the spin-average to lie in $I$. Recall that in order to obtain the mean field theory distribution we have to extremize the Gibbs functional given by
\begin{equation}
 \Phi[m]=H[m]-\frac{1}{\beta}S[m],
\end{equation}
where $\beta=T^{-1}$, $H[m]$ is the energy and $S[m]$ is the entropy functional. The latter can be written as:
\begin{equation}
S[m]=-\sum_{m_\beta=\{0,1\}} n_i P(m_\beta(i)) \log P(m_\beta(i)), 
\label{entropy}
\end{equation}
where $n_i$ is the degeneracy of the state. 

%%%%%%%%%%%%%%%%%%%%%%%%%%%%%%%%%%%%%%%%%%%%%%%%%%%%%

\subsection*{Case I: Non-degenerate edge states}

In this subsection we focus on the case in which the states \textit{on} and \textit{off} 
are not degenerate, so that $n_i=1$. In the next subsection we will deal with non-degenerate edge states and in particular with 3-degenerate \textit{on} states. 

In the process of extremizing the Gibbs functional we will see how the average valence of the graph naturally emerges. We impose:
$$\partial_{m_\beta} \Phi[m]=0. $$
Using
$$\partial_{m_\beta} S[m] = -\log(\frac{m_\beta}{1-m_\beta}) $$
and
$$ \partial_{m_\beta} H[m]=A\sum_{\alpha \in \mathscr B} J_{\alpha \beta} m_\alpha - B -C \sum_{\alpha \gamma \in \mathscr B, \alpha\neq\gamma\neq\beta} \tilde Q_{\alpha \beta \gamma} m_\alpha m_\gamma, $$
we find that the distribution for the $m_\alpha$ is
\begin{equation}
m_\beta = \frac{e^{-\beta \partial_{m_\beta} H[m]}}{1+e^{-\beta \partial_{m_\beta} H[m]}}= \frac{1}{1+e^{\beta \partial_{m_\beta} H[m]}}.
\label{gdist}
\end{equation}
The solution of this equation gives the equilibrium value of $m_\beta$ once the value of the temperature is fixed.

 We now want to write (\ref{gdist}) as a function of an average quantity on the graph. Let us first note that, in the mean field theory approximation, we have
\begin{equation}
 \sum_{\alpha} J_{\alpha \beta} m_\alpha = 2\ d (T),
\end{equation}
where $d(T)$ is the mean valence of the graph.  The valence $d(T)$ is a good graph observable that we can use also as a double check for our procedure since it appears explicitly in the original formulation of 
the hamiltonian and in the low temperature regime must take the value $v_0$. 
First, we note that:
$$
m_\alpha = \frac{N_{\text{on edges}}}{N(N-1)/2}=\frac{\sum_{i\in \mathscr I} d(i) /2} {N(N-1)/2}=\frac{d(T)}{N-1}. 
$$
In the first equality, $N_{\text{on edges}}$ is the number of edges of the graph which are in an \textit{on} state. In the second equality, the average valence (the sum over all the local valencies divided by the 
number of vertices) is explicitly written as a temperature dependent quantity.  In the third equality we used the graph property:
$$
\sum_{i \in \mathscr I} \frac{d(i)}{N-1} = \langle d \rangle \equiv d (T).
$$

The most complicated term in the hamiltonian is the 3-loop one. The simplest way to deal with it is to use the Ansatz dictated by the mean field theory:
\begin{equation}
\sum_{\alpha \gamma \in \mathscr B} \tilde Q_{\alpha \beta \gamma} m_\alpha m_\beta \approx \xi(T) d^2(T).
\end{equation}
Let us replace $m_\beta$ with its average value: $d(T)/{N-1}$.  Using eq.\  (\ref{property1a}) for non-retracing paths and assuming $N\gg1$ we obtain the dependence on $d(T)$. 
$\xi(T)$ is a function of order $\sim 1$ at low temperature, which we assume is dependent on $T$. Using these approximations we can see that $d(T)$ is a natural order parameter for our mean field theory
since it is easily recognized as implicitly defined in the stable distribution:
\begin{equation}
d(T) = \frac{N-1}{1+ e^{\beta [2 d(T) A-\xi \frac{C}{2} d^2(T) -B ]}}.
\label{eqdist}
\end{equation}

Again, in order to double check our procedure, we can ask if such an order parameter behaves as expected at low temperature. We must keep in mind that the starting hamiltonian (\ref{hamiltonian}) was constructed in such a 
way that the average valence at zero temperature was a fixed value of the parameter  $v_0$ at finite $N$.
We can now use (\ref{eqdist}) to check if this is the case. To do so, we Taylor expand both sides and match the zeroth and first order coefficients on the left and right hand side of the equation. That is, we start with the 
expansion
\begin{equation}
d(T)=\tilde \alpha + \tilde \beta T + O(T^2),
\label{dexp} 
\end{equation}
and, for the approximation to be consistent at $T=0$, we require  analyticity of the order parameter (this has to be the case for a finite volume system in ordinary statistical mechanics, which is the case for 
finite  $N$) . We  then require that inside the exponential of equation (\ref{eqdist}) the temperature independent terms 
in the numerator cancel out so that at $T=0$ the exponent is well defined.  This gives the second order equation in $\alpha$:
$$ 
2 \alpha A- \xi(0) C \alpha^2 = B.
$$
Now note that, while this equation has two solutions, we need to only look for the one which is analytical in the parameters of the model and  tends smoothly to the solution $\tilde \alpha=\frac{B}{2A}$ in the $C\rightarrow0$ limit. 
This fixes $\alpha$ to the value $\tilde \alpha$, given by
\begin{equation}
\tilde \alpha = \frac{A}{\xi(0) C}\left(1-\sqrt{1-\frac{C\xi(0) B}{A^2}}\right).
\label{cond1}
\end{equation}
We can now plug $\tilde \alpha$ at $T=0$ into (\ref{eqdist}):
$$
\tilde \alpha = \frac{N-1}{1+e^{(2A-\xi(0) C\tilde \alpha)\tilde \beta} },
$$
to obtain the value of $\tilde \beta$ in (\ref{dexp}):
\begin{equation}
\tilde \beta = \frac{1}{2A- \xi(0) C\tilde \alpha} \log\left(\frac{N-1}{\tilde \alpha} -1\right).
\label{cond2}
\end{equation}

It is easy to see that in the limit $N\rightarrow \infty $ we have $ \tilde \beta \rightarrow \infty$,
indicating a second-order phase transition (a discontinuity in the first derivative of the order parameter). In our case, this happens at $T=0$, meaning that this transition 
is not possible because there is no way to cool down the system to zero temperature with an external bath. However, we have to remember that we are just approximating the real system with a 
semi-classical analog. We then simply interpret the above result as the fact that the system reaches the ground state very quickly when the temperature approaches zero.

It is interesting now to plug in the couplings. Inserting equations (\ref{couplings}) into (\ref{cond1}), we find that at $T=0$
\begin{equation}
d(T=0)=\tilde \alpha = \frac{A}{\xi C}\left(1-\sqrt{1-\frac{C\xi B}{A^2}}\right) = \frac{6 pg_V}{\xi(0) g_L r^3}\left(1-\sqrt{1-\xi(0)\frac{g_L r^3 v_0}{3 p g_V}}\right).
\end{equation}
Note that, for small values of $r$, when $r^3\ll \frac{3 p g_V}{g_L v_0}$, we have $\tilde \alpha=v_0$,  meaning that at low temperature the mean degree is the one imposed by the degree term of the hamiltonian, as
expected. We can, however, see how the 3-loops term contributes to this quantity by a Taylor expansion in $r$:
\begin{equation}
d(T=0)=\tilde \alpha = v_0\left(1+\frac{2}{3}\xi(0) \frac{g_L r^3 v_0}{p g_V}\right).
\end{equation}

From this expression it is clear that the loop terms are suppressed if $g_V\gg g_L$.  This is the main result derived in this paper using the line graph representation. 
A plot of the function $d(T)$ is shown in Figure 3.  
\begin{figure}
 \includegraphics[scale=0.6]{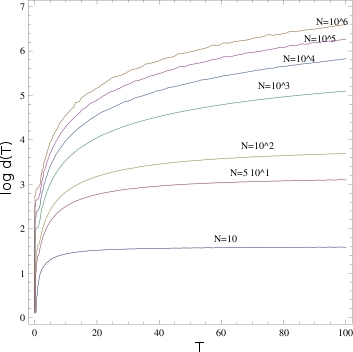}
\caption{The behavior of log($d(T)$) ($y$-axis) against $T$ ($x$-axis), for increasing $N$. }
\end{figure}

We now have the tools to calculate the susceptibility function for the theory in the mean field theory approximation. 
Recall that the susceptibility function tells us how the system reacts to a variation of the external magnetic field. In our case, the magnetic field is the combination $2 g_V p v_0$ and we note that the parameter $v_0$ in the 
hamiltonian appears only here. We have the following
analogy: $v_0$ represents the external magnetic field, while $2 g_V p$ represents the spin-coupling combination.

In order to calculate the susceptibility, we assume that the constant B is site-dependent (i.e.\ a field). Thus, we have
\begin{equation}
\langle G_{\alpha \beta}\rangle_B=-\frac{1}{\beta} \frac{\partial}{\partial B_\beta}\frac{\partial}{\partial B_\alpha} F[h]=\frac{1}{\beta}\frac{\partial m_\beta}{\partial B_\alpha}.
\end{equation}
In particular, we are interested in the susceptibility function when $T\approx 0$. From the study of it we can gain some information about the
low energy behavior of the model. We expand the equilibrium distribution
\begin{equation}
 m_\beta = \frac{1}{2}\left[1-\beta\left(A \sum_{\alpha \in \mathscr B} J_{\alpha \beta} m_\alpha-B_\beta-C \sum_{\alpha \gamma \in \mathscr B}\tilde Q_{\alpha \beta \gamma} m_\alpha m_\gamma\right)\right].
\end{equation}
Using the notation $\rho_{\alpha \beta}:=\sum_{\gamma \in \mathscr B}\tilde Q_{\alpha \beta \gamma}  m_\gamma$, we obtain
\begin{equation}
\tilde B_\beta = B_\beta-\frac{1}{2}=\sum_{\alpha \in \mathscr B}\left(2 \frac{\delta_{\alpha \beta}}{\beta}+ A J_{\alpha \beta}+ C \tilde \rho_{\alpha \beta}\right) m_{\alpha}.
\label{toinvert} 
\end{equation}
To invert this equation, we approximate $\rho_{\alpha \beta}$ by replacing $m_\gamma \rightarrow d(T)/(N-1)$:
$$
\sum_{\gamma \in \mathscr B}\tilde Q_{\alpha \beta \gamma}  m_\gamma\rightarrow \frac{d(T)}{N-1} \sum_{\gamma \in \mathscr B} \tilde Q_{\alpha \beta \gamma }.
$$
We can now can use property (\ref{property2b}) of the $\tilde Q$ matrices to find that the sum of the $\tilde Q$'s reduces to the incidence matrix of $\mathscr L(K_N)$.
Hence,  inverting  equation (\ref{toinvert}), we obtain
$$
m_\beta=\sum_\gamma Q_{\gamma \beta} \tilde B_\gamma,
$$
where $Q_{\gamma \beta}={(2 \frac{\delta}{\beta}+c_0 \textbf{J})^{-1}}_{\gamma \beta} $  and $c_0$ is an effective constant in front of the Ising term of the hamiltonian:
\begin{equation}
c_0 \approx p g_V + g_L \frac{r^3}{3!}\frac{d(T)}{N-1}.
\end{equation}
It is interesting to note that, 
thanks to property (\ref{property2a}), we can sum all the loop terms up to a finite number $1\ll\tilde L\ll N$ in the hamiltonian if we assume the mean field theory approximation. Inserting the couplings, we find
\begin{equation}
 c_0 \approx p g_V + g_L \sum_{L=3}^{\tilde L} \frac{r^2}{L!}\left(r\frac{d(T)}{N}\right)^{L-2} N^{L-3} \approx p g_V  + \frac{g_L r^2}{N d(T)^2}\left(e^{r\ d(T)}-1-r\ d(T)-\frac{r^2}{2}d(T)^2\right).
\label{effective2}
\end{equation}
It is interesting to note that in the limit $r\rightarrow 0$, or $T\rightarrow 0$ (where $d(T)$ tends to a finite number for $N\gg1$), this effective
constant tends to $p g_V$. We interpret this as the fact that at low temperature the loops become less and less important and the model is dominated 
by the Ising term. In particular, since the external ``magnetic'' field is given by $v_0$ and is assumed to be nonzero, it is not surprising that at $T=0$ the average valence, the equivalent ``magnetization'', 
approaches this value. We note that the $N\rightarrow \infty$ limit does not behave well unless $L=3$.  Higher loops are highly non-local objects.   For a given pair of  edges, all the $L$-loops based on these edges span 
the whole graph already at $L=4$, while of course this is not the case for $3$-loops.  As a result, in formula (\ref{effective2})  there is a factor proportional to $N^{L-3}$ which	 is not present at $L=3$.

%%%%%%%%%%%%%%%%%%%%%%%%%%%%%%%%%%%%%%%%%%%%%%%%%%%%%

\subsection*{Case II:  Degenerate edge states}

The Quantum Graphity model \cite{graphity1} allows for degenerate {\em on} states on the edges or the vertices of the graph.  Degeneracy of edge states is necessary, for instance, in order to have emergent matter via the 
string-net condensation mechanism of Levin and Wen.  Degeneracy requires modifying our calculations above and we will address it in this subsection.

The first possible generalization of the Quantum Graphity model is to introduce a Hilbert space on the edges of the form (\ref{fermhilb}):
$$
\mathscr H^{e}_{\beta}=\text{span}\{|0\rangle_{\beta},|1_1\rangle_{\beta},|1_2\rangle_{\beta},|1_3\rangle_{\beta}\}.
$$
This changes the degeneracy number in  equation (\ref{entropy}). With $n_1=3$ and $n_0=1$, we obtain
$$
\partial_{m_\beta} S[m]=-\log(\frac{m_\beta^3}{1-m_\beta})-2.
$$
The equilibrium distribution solves this equation. If we put $Q=\exp[\beta(\partial_{m_\beta}H[m]-2)]$, we have
\begin{equation}
 m_\beta = \left(\frac{2}{3}\right)^{\frac{2}{3}} \frac{Q}{\left(9+\sqrt{3}\sqrt{27+4 Q^3}\right)^{\frac{2}{3}}}+\frac{\left(9+\sqrt{3}\sqrt{27+4 Q^3}\right)^{\frac{1}{3}}}{2^{\frac{1}{3}} 3^{\frac{2}{3}}},
\label{distext}
\end{equation}
obtained from the only real solution of the third order polynomial equation $m_\beta^3+Q m_\beta-1=0$. 

Using the same procedure as before, it is easy to see that
equation (\ref{cond1}) remains unchanged: the low energy average valence is the same in both cases. However, the first derivative, that is, the coefficient of the $T$ term in
the Taylor expansion of the average valence in the temperature, changes, so that $\tilde \beta_{3,1}\geq\tilde \beta_{1,1}$ (with the obvious notation for the two coefficients).
This phenomenon can be understood using the following  argument. At high temperature, the two models behave in the same way, forcing the valence to be high. When the temperature
drops, $d(T)$ also goes down. While in the (1,1) case the phase space of the \textit{on} edges is the same as that of the \textit{off} edges, in the (3,1) case the system prefers to
stay in the \textit{on} state. Thus, when the temperature decreases the system (3,1) is, at first, slowly converging to the ground state, but at $T=0$ it is forced to go to the ground 
state. For this reason the function $d(T)$ has a greater derivative near $T=0$ in the (3,1) case.

%%%%%%%%%%%%%%%%%%%%%%%%%%%%%%%%%%%%%%%%%%%%%%%%%%%%%

\section{Conclusions}
In this paper we introduced a technique to map the Quantum Graphity hamiltonian on the line graph of a complete graph. 
This procedure requires the introduction of the Kirchhoff matrix of a graph and the $n$-string matrices related to these. 
This mapping is general and not specific to Quantum Graphity. Using this mapping in a weak coupling approximation of 
the model, the mean field theory approximation and the low temperature expansion, we studied 
the properties of the model near zero temperature after having identified the average degree as a order parameter.
We found that the model is dual to an Ising model with external nonzero magnetic field if we neglect the interaction terms due to 
loops.
In particular, we showed that the average valence is naturally a good order parameter for the mean field theory approximation and 
we found, implicitly, its average distribution using the mapped hamiltonian and the mean field
theory approximation for the 3-loop term. We
then studied the susceptibility function and showed how the duality with the Ising model can help to interpret the results. 
In particular,  the parameter $v_0$ plays the role of the external magnetic field. Since $v_0$ is assumed 
to be never zero,
the model has no phase transition and at $T=0$ the system goes to the ground state as expected. In fact, we found that at zero 
temperature the mean valence is determined by the parameter $v_0$ and we approximated
the first order correction, showing the dependence on the coupling constants of the model. While these results were expected on 
general grounds, the mapping used here simplified the problem and allowed a quantitative analysis.

What emerged from the study of the average distribution for the valence is that, if the vertex valence term dominates ($g_{V}\gg g_{L}$), 
the loop term corrections to the average valence of the ground state are suppressed at 
$T=0$.  We found the dependence on the coupling constants explicitly. 
This result is confirmed by the study of the susceptibility. Thanks to the mean field theory approximation, we found the contribution 
of all  loops to the susceptibility and showed that the susceptibility function tends to the 
Ising one when $T$ approaches zero.
We then applied this  procedure to the degenerate case, and showed that the degeneracy does not change the average valence at low 
temperature but only the speed with which this ground state is reached. As a final remark,
we stress that the vertex valence is an important quantity in the model. 
In fact, as shown in \cite{iaif} using the Lieb-Robinson bound, 
the speed with which information can propagate on graphs is bounded 
by a valence-dependent quantity. For this reason, as the temperature drops, the speed of the emergent light field must drop 
with the valence.

We would like to stress that we assumed that the system was at equilibrium with an external bath. The problem of the required 
external bath in the model has been studied in \cite{graphity2}, where additional degrees of freedom (bosonic particles on vertices) 
were introduced  to let the system thermalize and reach an equilibrium distribution. 

%%%%%%%%%%%%%%%%%%%%%%%%%%%%%%%%%%%%%%%%%%%%%%%%%%%%%

\section{Aknowledgements}
The authors are indebted to Cohl Furey, Piero Porta Mana, Isabeau Pr\'{e}mont-Schwarz, Simone Severini,  Lee Smolin, and especially Alioscia Hamma
for reading the manuscript and providing useful advice and comments. This work was supported by NSERC grant RGPIN-312738-2007 and the Humboldt Foundation.  
Research at Perimeter Institute is supported by the Government of Canada through Industry Canada and by the Province of Ontario
through the Ministry of Research \& Innovation.

%%%%%%%%%%%%%%%%%%%%%%%%%%%%%%%%%%%%%%%%%%%%%%%%%%%%%

\end{document}